\documentclass[reprint,aps,prl,superscriptaddress,longbibliography]{revtex4-1}
\usepackage[latin9]{inputenc}
\usepackage{color}
\usepackage{times}
\usepackage{amsmath}
\usepackage{amssymb}
\usepackage{stmaryrd}
\usepackage{graphicx}
\usepackage[unicode=true,
 bookmarks=true,bookmarksnumbered=false,bookmarksopen=false,
 breaklinks=false,pdfborder={0 0 1},backref=false,colorlinks=true]
 {hyperref}
\hypersetup{
 linkcolor=magenta, urlcolor=blue, citecolor=blue, pdfstartview={FitH}, hyperfootnotes=true, unicode=true}

\makeatletter

\@ifundefined{textcolor}{}
{
 \definecolor{BLACK}{gray}{0}
 \definecolor{WHITE}{gray}{1}
 \definecolor{RED}{rgb}{1,0,0}
 \definecolor{GREEN}{rgb}{0,1,0}
 \definecolor{BLUE}{rgb}{0,0,1}
 \definecolor{CYAN}{cmyk}{1,0,0,0}
 \definecolor{MAGENTA}{cmyk}{0,1,0,0}
 \definecolor{YELLOW}{cmyk}{0,0,1,0}
}

\usepackage{amsfonts}\usepackage{tabularx}\usepackage{dcolumn}\usepackage{bm}\usepackage{graphicx}\usepackage{epstopdf}

\hypersetup{urlcolor=blue}

\usepackage{tensor}
\usepackage{braket}

\DeclareMathOperator{\sgn}{sgn}
\newcommand{\Hc}{\mathrm{H.c.}}
\renewcommand{\(}{\left(}
\renewcommand{\)}{\right)}

\DeclareMathOperator{\tr}{tr}

\usepackage[capitalise,compress]{cleveref}
\crefname{section}{Sec.}{Secs.}
\Crefname{section}{Section}{Sections}
\crefrangelabelformat{equation}{\textup{(#3#1#4)}--\textup{(#5#2#6)}}

\makeatother
\begin{document}

\title{Asymmetric Particle Transport and Light-Cone Dynamics Induced by Anyonic Statistics}

\author{Fangli Liu}
\affiliation{Joint Quantum Institute, NIST/University of Maryland, College Park, Maryland 20742, USA}

\author{James R. Garrison}
\affiliation{Joint Quantum Institute, NIST/University of Maryland, College Park, Maryland 20742, USA}
\affiliation{Joint Center for Quantum Information and Computer Science, NIST/University of Maryland, College Park, Maryland 20742, USA}

\author{Dong-Ling Deng}
\affiliation{Center for Quantum Information, IIIS, Tsinghua University, Beijing 100084, PR China}
\affiliation{Condensed Matter Theory Center, Department of Physics, University of Maryland, College Park, Maryland 20742, USA}
\affiliation{Joint Quantum Institute, NIST/University of Maryland, College Park, Maryland 20742, USA}

\author{Zhe-Xuan Gong}
\affiliation{Department of Physics, Colorado School of Mines, Golden, Colorado 80401, USA}
\affiliation{Joint Quantum Institute, NIST/University of Maryland, College Park, Maryland 20742, USA}
\affiliation{Joint Center for Quantum Information and Computer Science, NIST/University of Maryland, College Park, Maryland 20742, USA}

\author{Alexey V. Gorshkov}
\affiliation{Joint Quantum Institute, NIST/University of Maryland, College Park, Maryland 20742, USA}
\affiliation{Joint Center for Quantum Information and Computer Science, NIST/University of Maryland, College Park, Maryland 20742, USA}

\begin{abstract}
We study the non-equilibrium dynamics of Abelian anyons in a one-dimensional system. We find that the interplay of anyonic statistics and interactions gives rise to spatially asymmetric particle transport together with a novel dynamical symmetry that depends on the anyonic statistical angle and the sign of interactions. Moreover, we show that anyonic statistics induces asymmetric spreading of quantum information, characterized by asymmetric light cones of out-of-time-ordered correlators. Such asymmetric dynamics is in sharp contrast with the dynamics of conventional fermions or bosons, where both the transport and information dynamics are spatially symmetric. We further discuss experiments with cold atoms where the predicted phenomena can be observed using state-of-the-art technologies. Our results pave the way toward experimentally probing anyonic statistics through non-equilibrium dynamics.
\end{abstract}


\maketitle

Fundamental particles in nature can be classified as either bosons or fermions, depending on their exchange statistics.
However, other types of quantum statistics are possible in certain circumstances.
For instance, Abelian anyons are characterized by fractional statistics interpolating
between bosons and fermions~\cite{Leinaas1977, goldin1981representations, Frank84, Tsui82, Laughlin83}. When two anyons are exchanged,  their joint wavefunction picks up a generic phase factor, $e^{i\theta}$. Anyons play important roles in several areas of modern physics research, such as fractional quantum Hall systems~\cite{Laughlin83, Halperin84, Arovas} and spin liquids~\cite{KITAEV20062, Yao07, Bauer2014}, not only because of their fundamental physical interest, but also due to their potential applications in topological quantum computation and information processing~\cite{Kitaev03, Sarma05, Bonderson06, Stern06, Nayak08, Alicea11, Stern13}.   In the beginning, the exploration of anyons was restricted to two-dimensional systems. Later, Haldane generalized the concept of fractional statistics and anyons to arbitrary dimensions~\cite{Haldane91, Haldane912}.

The physics of Abelian anyons in one dimension (1D) has attracted a great deal of recent interest~\cite{Ha94, Murthy94, Wu95, Amico98, Mazza, Zinner15, Kundu99, Batchelor06, Girardeau06, del08, Calabrese07, Zatloukal2014, Greiter2009, Hao08, Hao09, Tang15, Hao12}.
Anyons in 1D exhibit a number of intriguing
properties, including statistics-induced quantum phase transitions~\cite{Keilmann11, Greschner15, SilvaValencia2016, SilvaValencia2018}, asymmetric momentum distribution in ground states~\cite{Keilmann11, Lange17, Hao08, Hao09, Hao12, Tang15, Greiter2009}, continuous fermionization of bosonic atoms~\cite{Strater16}, and anyonic symmetry protected topological phases~\cite{Lange17}.
Several schemes have been proposed for implementing anyonic statistics in ultracold atoms~\cite{Keilmann11, Greschner15, Strater16, Lange17, Clark18} and photonic systems~\cite{Fan17} by engineering occupation-number dependent hopping using Raman-assisted tunneling~\cite{Keilmann11, Greschner15} or periodic driving~\cite{Strater16, Fan17}.
Cold atom quantum systems~\cite{Greiner02, Lewen07, Bloch08} are powerful platforms not only for probing equilibrium properties of many-body systems, but also for studying uncharted  non-equilibrium physics~\cite{Eisert15, Gogolin16, Schnei12, Ronzh13, Pertot14, Hung13, Kaufman16, Zhang17, Flsch16, Jurce17}. 
Yet, most of the non-equilibrium studies to date have focused on fermionic or bosonic systems, where anyonic statistics do not come into play. 

In this work, we study the interplay between anyonic statistics and non-equilibrium dynamics.
In particular, we study the particle transport and information dynamics of Abelian anyons in 1D, motivated by recent proposals~\cite{Keilmann11, Greschner15, Strater16, Lange17} and the experimental realization of density-dependent tunneling~\cite{Clark18,Meinert16}, as well as by technological advances in probing non-equilibrium dynamics in ultracold atomic systems~\cite{Schnei12,Ronzh13}.
As we shall see, statistics plays an important role in the non-equilibrium dynamics of anyons.
First, distinct from the bosonic and fermionic cases, anyons in 1D exhibit \emph{asymmetric} density expansion under time evolution of a homogeneous anyon-Hubbard model (AHM\@).
The asymmetric transport is controlled by the anyonic statistical angle $\theta$ and interaction strength $U$. When the sign of $\theta$ or $U$ is reversed, the expansion changes its preferred direction, thus revealing a novel dynamical symmetry of the underlying AHM\@.
We identify this symmetry operator and analyze the asymmetric expansion dynamics using perturbation theory, confirming the important role played by statistics and interactions. 
In addition, we use the so-called out-of-time-ordered correlator (OTOC)~\cite{Larkin69} to characterize the spreading of information in such systems.
We find that information spreads with different velocities in the left and right directions, forming an asymmetric light cone.

In contrast to previous studies on ground-state properties~\cite{Calabrese07, Hao08, Hao09, Tang15, Keilmann11, Greschner15, Strater16, Lange17} or hard-core cases~\cite{del08, Hao12, Wright14} of 1D anyons, here we focus on the out-of-equilibrium physics of anyonic systems which can be implemented in experiment~\cite{Keilmann11, Greschner15, Strater16, Lange17, Clark18}.
Moreover, we focus mainly on observables that both reveal anyonic properties directly and can be probed in cold atom systems, where the anyonic statistics can be realized via correlated-tunneling terms~\cite{Strater16}.
Crucially, our work provides a new method for detecting anyonic statistics even in systems where the ground state is difficult to prepare.

\begin{figure*}
  \centering\includegraphics[width=\textwidth, height=6.1cm]{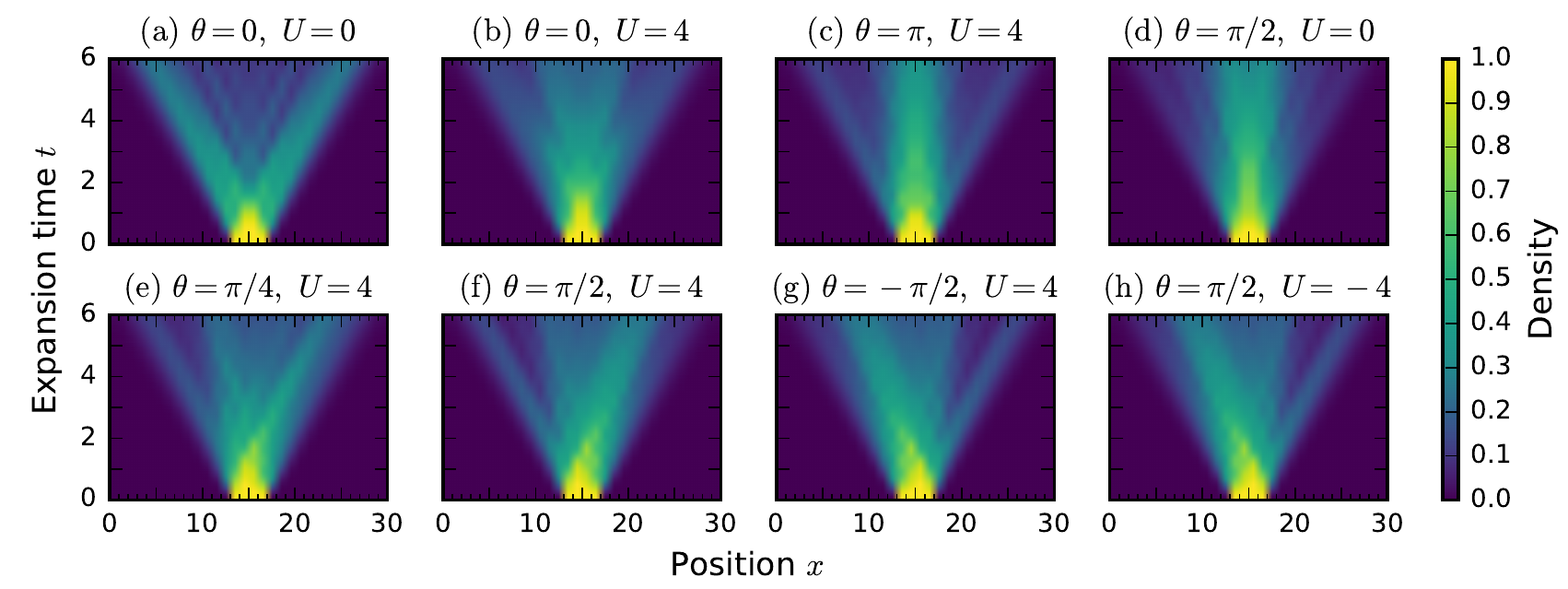}
  \caption{Density expansion dynamics for particles initially localized one-per-site in the central $N$ sites, with different statistical angles $\theta$ and interaction strengths $U$. In all plots, the particle number is $N=4$ and the lattice size is $L=30$. (a)--(b) Bosonic cases with zero  and non-zero interactions, respectively. (c) ``Pseudofermionic'' case ($\theta =\pi$) with non-zero interactions. (d)--(h) Anyonic cases with various values for $\theta$ and $U$.}
  \label{fig1}
\end{figure*}

{\it Model.---}%
We consider 1D lattice anyons with on-site interactions---the anyon-Hubbard model~\cite{Keilmann11, Greschner15, Strater16, Lange17, Clark18, Fan17}:
\begin{equation}
\hat{H}_A= -J\sum_{j=1}^{L-1} \( \hat{a}_j^{\dagger} \hat{a}_{j+1}^{\phantom\dagger} + \Hc \) + \frac{U}{2} \sum_{j=1}^{L} \hat{n}_j (\hat{n}_j-1),
\label{ahm}
\end{equation} 
where $\hat{n}_j= \hat{a}_j^{\dagger} \hat{a}_j^{\phantom\dagger}$, and $J$ and $U$ describe nearest-neighbor tunneling and on-site interaction, respectively. Throughout the paper, we set $J=1$ as the energy unit. The anyon creation ($\hat{a}_j^{\dagger}$) and annihilation
($\hat{a}_j$) operators obey the generalized commutation relations
\begin{alignat}{3}
& \left[ \hat{a}_j^{\phantom\dagger}, \hat{a}_k^{\phantom\dagger} \right]_{\theta} &&\equiv\hat{a}_j \hat{a}_k -e^{-i\theta \sgn(j-k)}\hat{a}_k \hat{a}_j &&= 0, \label{s2} \\
& \left[ \hat{a}_j^{\phantom\dagger}, \hat{a}_k^{\dagger} \right]_{-\theta} &&\equiv\hat{a}_j^{\phantom\dagger} \hat{a}_k^{\dagger}- e^{i\theta \sgn(j-k)}\hat{a}_k^{\dagger} \hat{a}_j^{\phantom\dagger} &&= \delta_{jk}, \label{s1}
\end{alignat}
where $\theta$ is the anyonic statistical angle. Here, $\sgn(k) =-1, 0, 1$ for $k<0$, $=0$, $>0$, respectively. \Cref{s2,s1} imply that
particles on the same site behave as bosons. When $\theta= \pi$, these lattice anyons are ``pseudofermions,'' as they behave like fermions off-site, while being bosons on-site~\cite{Keilmann11}.

By a generalized, fractional Jordan-Wigner transformation,
$\hat{a}_j= \hat{b}_j e^{-i\theta \sum_{k=1}^{j-1}\hat{n}_k}$,
the above AHM can be mapped to an extended Bose-Hubbard model (EBHM),
\begin{equation} \label{eq:EBHM}
\hat{H}_B=  -J\sum_{j=1}^{L-1} \( \hat{b}_j^{\dagger} \hat{b}_{j+1}^{\phantom\dagger} e^{-i\theta \hat{n}_j} + \Hc \) + \frac{U}{2} \sum_{j=1}^{L} \hat{n}_j (\hat{n}_j-1) ,
\end{equation}
where $\hat{b}_j$ is the bosonic annihilation operator for site $j$, and $\hat n_j = \hat a^\dagger_j \hat a^{\phantom\dagger}_j = \hat b^\dagger_j \hat b^{\phantom\dagger}_j$~\cite{Kundu99, Batchelor06, Girardeau06, Keilmann11, Greschner15, Strater16}. Under this transformation, anyonic statistics have been translated to density-dependent hopping terms, which are the key ingredient to implementing anyonic statistics in 1D\@.  As mentioned,  one can realize such terms in cold atomic systems using either Raman-assisted tunneling~\cite{Keilmann11, Greschner15} or time-periodic driving~\cite{Strater16, Fan17, Clark18}. 

{\it Asymmetric particle transport.---}%
We consider the expansion dynamics of anyons initially localized at the central region of a 1D lattice, one per occupied site. The initial state can be written as a product state in Fock space, $\ket{\Psi_0}_A= \prod_i \hat{a}_i^{\dagger} \ket{0}$, with occupied sites distributed symmetrically around the lattice center. At times $t>0$, the system evolves under $\hat{H}_A$ [\cref{ahm}].  This procedure is equivalent to a quantum quench from $U/J= \infty$ to finite   $U/J$. To characterize particle transport, we study  the dynamics of the \emph{real} space anyon density,
$n^A_j(t) = \tensor*[_A]{\bra{\Psi_0} e^{i\hat{H}_At} \hat{n}_j e^{-i\hat{H}_A t} \ket{\Psi_0}}{_A}$,
where we have set $\hbar=1$. Under the fractional Jordan-Wigner transformation, the particle number operator $\hat{n}_j$ remains invariant (i.e.\ $\hat a^\dagger_j \hat a^{\phantom\dagger}_j = \hat b^\dagger_j \hat b^{\phantom\dagger}_j$), $\hat{H}_A$ maps to $\hat{H}_B$, and the initial state picks up an unimportant phase $\phi$, i.e.\ $\ket{\Psi_0}_A= e^{i\phi}\prod_i \hat{b}_i^{\dagger} \ket{0}=e^{i\phi} \ket{\Psi_0}_B$.
These relations directly lead to the following equality:
\begin{equation}
n_j^A(t) = \tensor*[_B]{\bra{\Psi_0} e^{i\hat{H}_Bt} \hat{n}_je^{-i\hat{H}_B t} \ket{\Psi_0}}{_B} = n_j^B(t), \label{anybos}
\end{equation}
which indicates that anyonic and bosonic particle densities are equivalent under time evolution governed by their respective initial states and Hamiltonians. \Cref{anybos} maps anyonic density to  bosonic density, which can be directly measured in cold atom experiments~\cite{Keilmann11, Greschner15, Strater16, Lange17, Schnei12, Ronzh13}. 
Likewise, the state $\ket{\Psi_0}_B$ can be easily prepared in such experiments~\cite{Schnei12, Ronzh13}.

Exact diagonalization results on the expansion dynamics for a variety of  statistical angles and interaction strengths are shown in Fig.~\ref{fig1}.
Figures~\ref{fig1}(a) and (b) show transport dynamics for the bosonic case ($\theta=0$). Consistent with experimental observations in Ref.~\cite{Ronzh13}, bosons exhibit ballistic expansion when $U=0$ [Fig.~\ref{fig1}(a)]. However, any finite interaction strength ($U\ne 0$) breaks the integrability of the Bose-Hubbard model and dramatically suppresses the density expansion [Fig.~\ref{fig1}(b)], leading to diffusive (i.e., non-ballistic) dynamics~\cite{Ronzh13}.  In contrast to bosonic cases, for anyons  with non-zero $\theta$ and even \emph{vanishing} interaction strength, the transport shows strong signatures of being diffusive rather than ballistic [see Fig.~\ref{fig1}(d)]. This implies that  anyonic statistics itself can break integrability and act as a form of effective interaction~\cite{Moramp17}, as is immediately clear from the correlated-tunneling terms in the EBHM in Eq.\ (\ref{eq:EBHM}). 
From Figs.~\ref{fig1}(a) and (d), we also note that for bosons or anyons with zero interaction strength, the density expansion is symmetric. 

\begin{figure}
  \centering\includegraphics[width=0.5\textwidth]{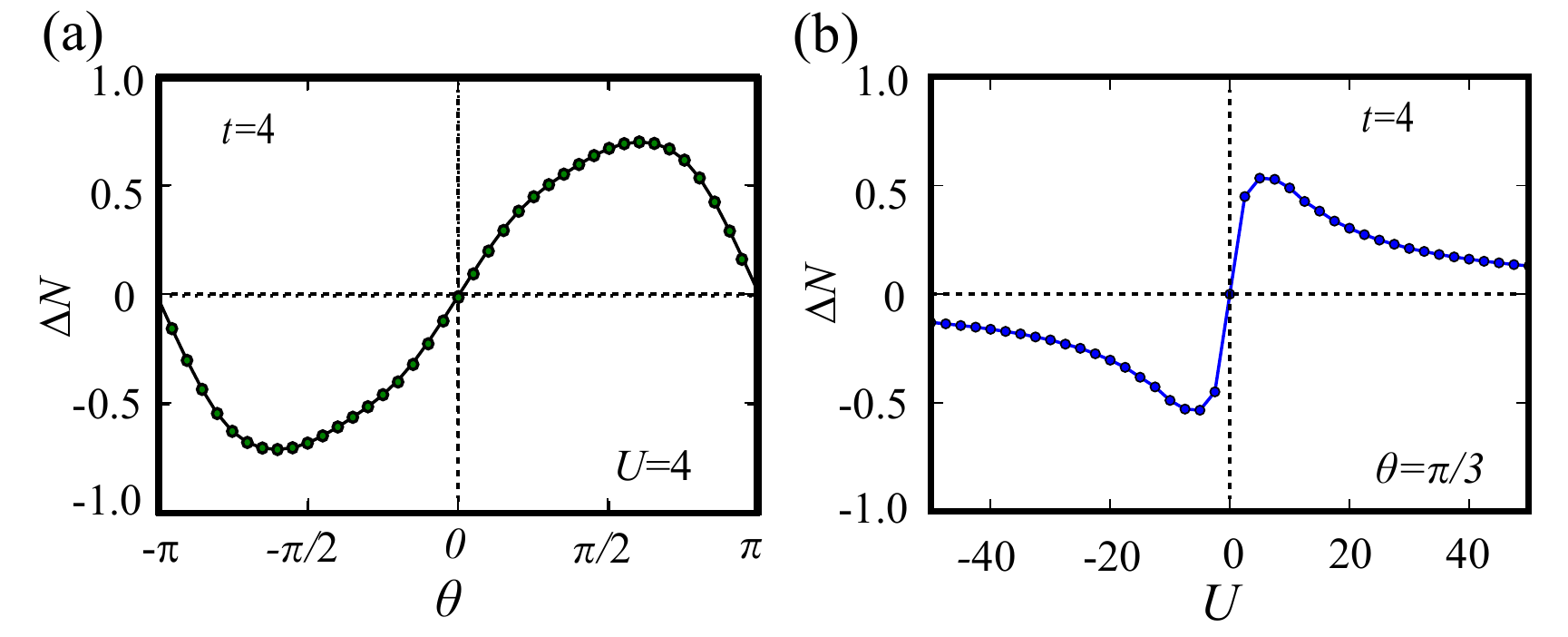}
  \caption{(a) Particle number difference $\Delta N$ between the right and left halves versus anyon angle $\theta$ at time $t=4$, which is beyond the perturbative regime yet occurs before the quench hits the boundary.  The interaction strength is $U=4$.  (b) $\Delta N$ versus interaction strength $U$ at time $t=4$, with $\theta= \pi/3$. The particle number is $N=4$, and the lattice size is $L=30$ for both plots, just as in Fig.~\ref{fig1}.
  }
  \label{fig2}
\end{figure}

Different from the above symmetric transport, for anyons with
$0 < \theta < \pi $  and finite interaction strength $U$, the dynamical density distribution is asymmetric, with one preferred propagation direction [Figs.~\ref{fig1}(e)--(h)]. This is the most striking feature of anyonic statistics' effects on  transport behavior. Such asymmetric expansion is due to inversion symmetry breaking of the AHM~\cite{Keilmann11, Frank90}, a direct consequence of the underlying 1D anyonic statistics [\cref{s2,s1}]. A perturbation analysis reveals the important role played by statistics and interactions (see Supplemental Material for details~\cite{supp}).
Our results illustrate that anyonic statistics  has clear signatures in non-equilibrium transport, which may aid in their detection.
Previous works have suggested detecting anyonic statistics via asymmetric momentum distributions in equilibrium ground states~\cite{Keilmann11, Hao08, Hao09, Hao12, Tang15, Greschner15, Strater16}, but ground states are often difficult to prepare experimentally.

Figure~\ref{fig2}(a) plots one measure of the above-mentioned asymmetry, the particle number difference $ \Delta N = \sum_{i=1}^{L/2}( n_{i+L/2} - n_{i} )$  between two halves versus statistical angle $\theta$. The results indeed show clear dependence on the statistical parameter $\theta$, thus demonstrating that one can detect the underlying anyonic statistics using expansion dynamics.
\Cref{fig2}(b) shows the dependence of $\Delta N$ on interaction strength for fixed statistical angle.
We note that the largest asymmetric measure $\Delta N$ occurs for intermediate values of $U$, as the expansion dynamics are symmetric at both $U=0$ (analyzed below) as well as in the limit of large $U$ (the hard-core case)~\cite{del08, Hao12, Wright14}.

{\it Symmetry analysis.---}%
Comparing Figs.~\ref{fig1}(g) and (h) to Fig.~\ref{fig1}(f), we can clearly see that by reversing the sign of the statistical angle $\theta$ or interaction strength $U$, anyons also reverse their preferred propagation direction. This dynamical symmetry
is further illustrated in Figs.~\ref{fig2}(a) and (b), which provide evidence that $\Delta N$ is indeed an odd function of $\theta$ and an odd function of $U$. The results differ from experimental findings for fermionic/bosonic gases~\cite{Schnei12, Ronzh13}, where density expansion dynamics are identical for $\pm U$ (further analyzed in a recent theoretical work, Ref.~\cite{Yu17}).

To understand the dynamical symmetry, we focus on the symmetry properties of the mapped EBHM for convenience.
$\hat{H}_B$ explicitly breaks inversion symmetry $\mathcal{I}$, as the phase of the correlated-tunneling term depends only on the occupation number of the left site (which becomes the right site under inversion).
It also breaks time-reversal symmetry, as $\mathcal{T} e^{-i\theta \hat{n}_j} \mathcal{T}^{-1}= e^{i\theta \hat{n}_j}$.
However, if we consider the number-dependent gauge transformation $\mathcal{R}= e^{-i\theta \sum_j \hat{n}_j(\hat{n}_j-1)/2 }$ and define a new symmetry operator $\mathcal{K}= \mathcal{R}\mathcal{I}\mathcal{T}$, $\hat{H}_B$ is invariant under $\mathcal{K}$~\cite{Lange17, supp}:
\begin{equation}
\mathcal{K} \hat{H}_B \mathcal{K}^{\dagger}= \hat{H}_B.
\label{k_symmetry}
\end{equation}
The transformed EBHMs with the opposite sign of interaction or statistical angle are related by the number parity operator $\mathcal{P}= e^{i\pi\sum_r \hat{n}_{2r+1}}$ or the time-reversal operator $\mathcal{T}$, respectively:
\begin{align}
\mathcal{P} \hat{H}_{B, +U} \mathcal{P}^{\dagger} &= -\hat{H}_{B, -U},
\label{p_symmetry}\\
\mathcal{T} \hat{H}_{B, +\theta} \mathcal{T}^{-1} &= \hat{H}_{B, -\theta}. 
\label{t_symmetry}
\end{align}
Using \cref{k_symmetry,p_symmetry,t_symmetry}, one can derive the following relations~\cite{supp}:
\begin{align}
\langle \hat{n}_j(t)\rangle_{+U} &= \langle \hat{n}_{j^{'}}(t)\rangle_{-U}, \label{eq:dynsym1} \\
\langle \hat{n}_j(t)\rangle_{+\theta} &= \langle \hat{n}_{j^{'}}(t)\rangle_{-\theta}, \label{eq:dynsym2}
\end{align}
where $\langle \cdot\rangle$ denotes the expectation value of a Heisenberg operator taken with respect to the initial state given above, and sites $j, j^{'}$ are related by the inversion operator $\mathcal{I}$. In fact, the above equations hold for a more general class of initial states (see Supplemental Material~\cite{supp}).
Therefore, in contrast to fermionic/bosonic gases~\cite{Yu17} (symmetric expansion), the above relations indicate that anyons flip their preferred expansion direction when one changes the sign of $U$ or $\theta$ in \cref{ahm}.  The above equalities also immediately imply  when $\theta= 0$ or $\pi$ (bosons or ``pseudofermions,'' respectively) or when $U=0$, the transport is symmetric [shown in Figs.~\ref{fig1}(a)--(d)], consistent with previous results for integrable systems~\cite{del08, Hao12, Wright14}.

{\it Information dynamics.---}%
The spreading of information in an interacting quantum many-body system has received tremendous interest~\cite{Lieb72,Lauchli08,Chene12,Bohrdt17,Eisert15,Shen17,Luitz17}. For conventional fermionic or bosonic systems with translation invariance, information spreading occurs in a spatially symmetric way~\cite{Bohrdt17, Lauchli08, Chene12}. However, as we demonstrate below, this is not generally the case for anyonic systems, where statistics can manifest itself in the information dynamics.

We diagnose information spreading by examining the OTOC, a quantity that has received a great deal of recent interest in studies of quantum scrambling~\cite{Shen17, Luitz17, Huang17, CurtVK17, Nahum17, He17, Fan16, Chen16, Swingle17, Leviatan17, Swingle18, zhang18, swingle1802}.
We define the anyonic OTOC as
$C_{jk}(t)= \Braket{\lvert \left[ \hat{a}_j(t), \hat{a}_k(0)\right]_{\theta} \rvert^2 }_\beta$.
Here, $\braket{\cdot}_\beta$ is taken with respect to the thermal ensemble $e^{-\beta \hat{H}_A} / \tr ( e^{-\beta \hat{H}_A} )$ with inverse temperature  $\beta$. The use of the generalized commutator defined by \cref{s2,s1} ensures that  $C_{jk}(t)$ vanishes at $t=0$. It then starts to grow when quantum information  propagates from site $k$ to site  $j$~\cite{Huang17, Shen17, Luitz17, Bohrdt17}.  We focus on the out-of-time-ordered part of the above commutator,
\begin{equation}
F_{jk}(t)= \Braket{\hat{a}_j^{\dagger}(t) \hat{a}_k^{\dagger}(0) \hat{a}^{\phantom\dagger}_j(t) \hat{a}^{\phantom\dagger}_k(0) }_\beta e^{i\theta \sgn(j-k)}.
\label{otoc}
\end{equation}

\begin{figure}
  \centering\includegraphics[width=0.5\textwidth, height=6.5cm]{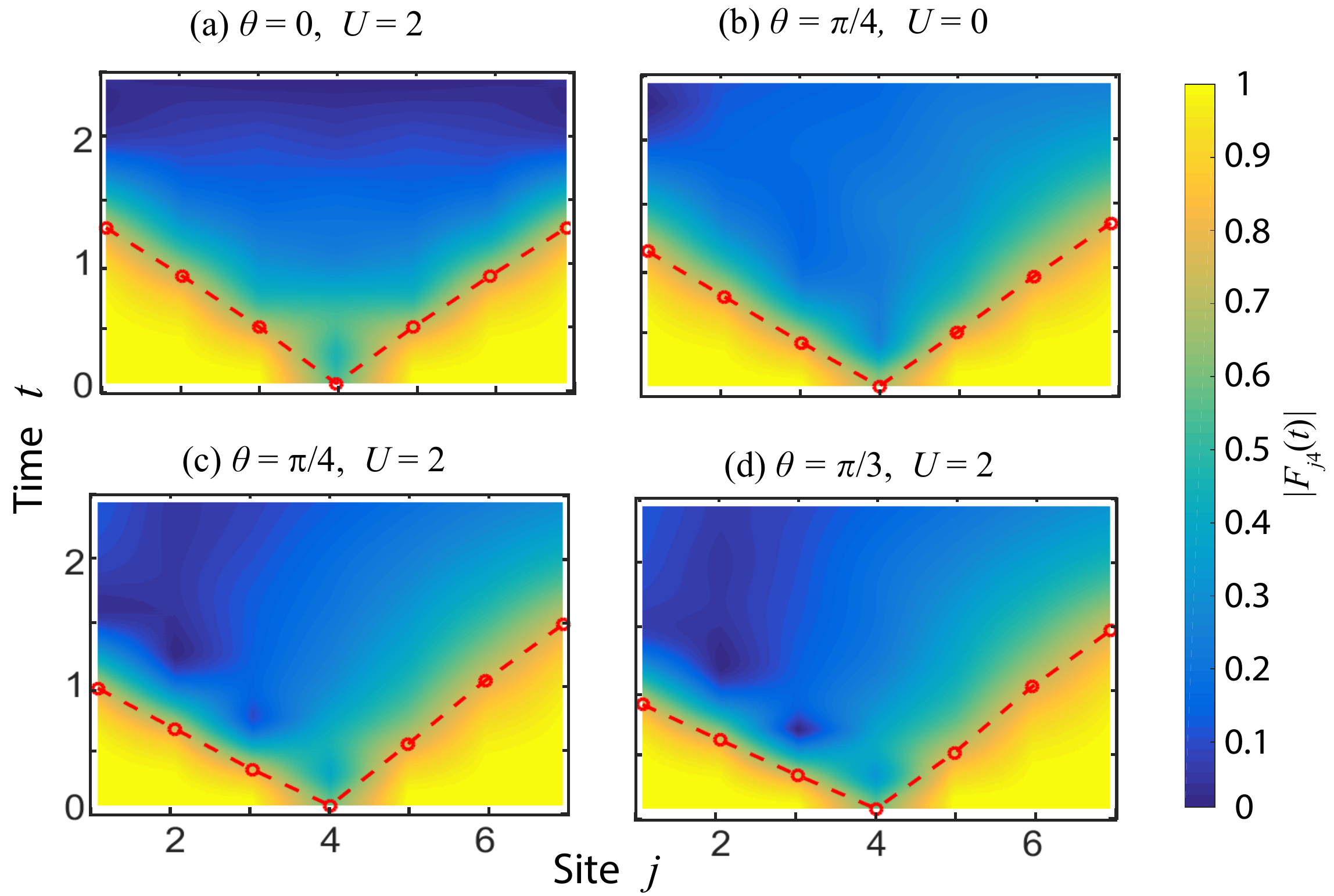}
  \caption{OTOC growth $|F_{jk}(t)|$ for different statistical angles $\theta$ and interaction strengths $U$. Here, $L=7$, $\beta^{-1}=6$, $k=4$, and the local Hilbert space of each site is truncated to three states.  Plotted is (a) a bosonic case ($\theta=0$) with non-zero interaction, as well as anyonic cases with (b) vanishing  and (c),(d) non-vanishing interaction strengths. The red dots denote where the OTOCs fall to 75\% of their initial values.  The colormaps are interpolated to non-integer $j$ to better illustrate the light cone behavior.
  }
  \label{fig3}
\end{figure}

Figures~\ref{fig3}(a)--(d) show numerical results for various interaction strengths $U$ and statistical angles $\theta$. In contrast to the density transport shown in Fig.~\ref{fig1}(b), quantum information spreads in a ballistic way for bosons even when $U \ne 0$~\cite{Chene12, Lauchli08}. Indeed, for bosons ($\theta=0$), the OTOCs map out a symmetric light cone, as shown in Fig.~\ref{fig3}(a). However, for  the anyonic case ($\theta\neq 0, \pi$), information propagation is asymmetric for the left and right directions [Figs.~\ref{fig3}(b)--(d)], resulting in an asymmetric light cone. We emphasize that this occurs even when $U=0$, as the aforementioned dynamical symmetry [\cref{eq:dynsym1,eq:dynsym2}] does not hold for the OTOC\@.

\begin{figure}
  \centering\includegraphics[width=0.48\textwidth, height=6.6cm]{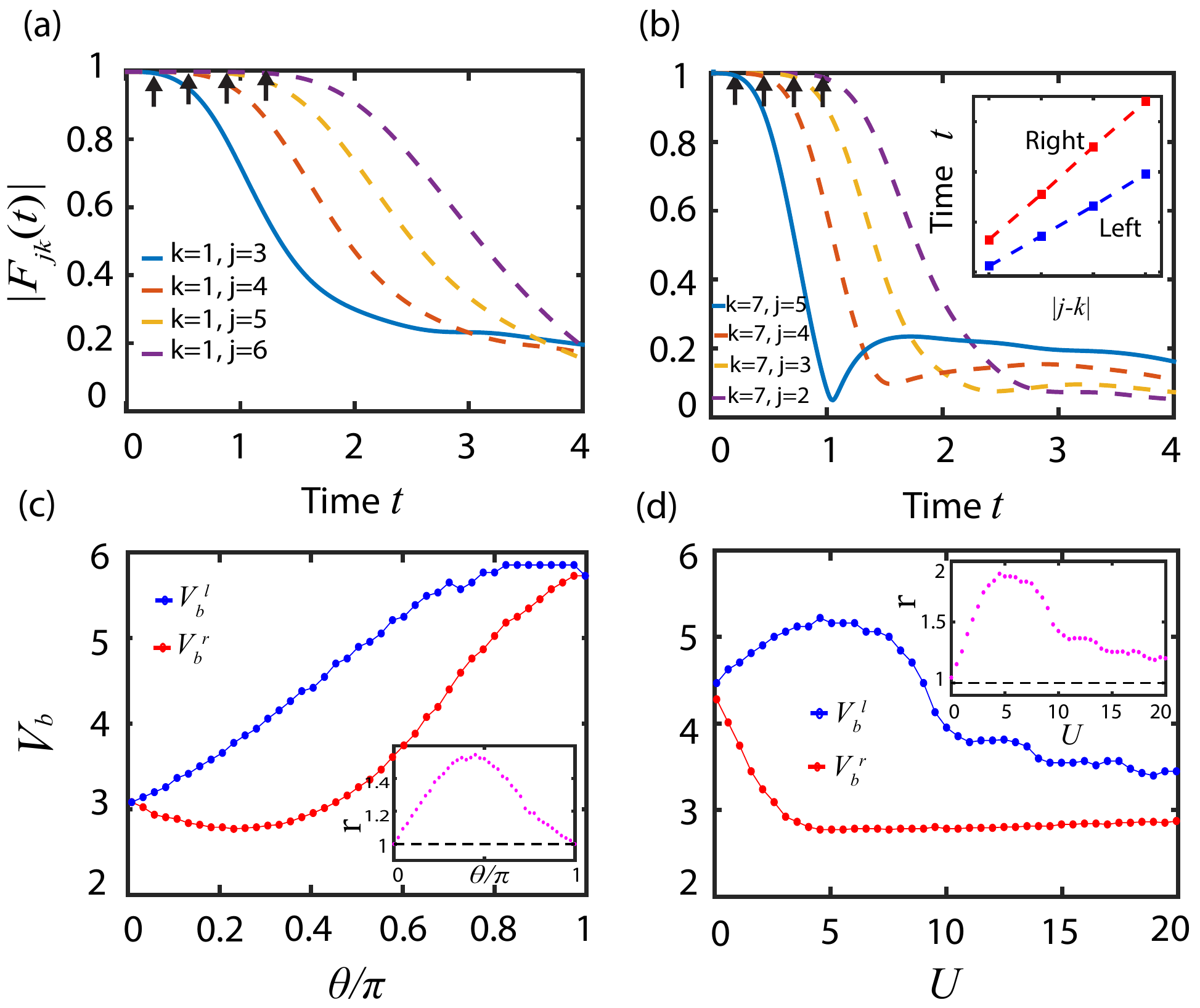}
  \caption{(a) OTOC growth characterizing quantum information spreading from the left-most site, $k=1$, rightward. The OTOC starts to fall when information reaches the $j$th site, and the black arrows denote the OTOCs' fall to 99\% of their initial values. Parameters: $\theta=\pi/3$, $U=2$, $L=7$. (b) Same as (a) but shows information spreading from the right-most site $k=7$ toward the left. Inset: linear fit to extract the butterfly velocities for left (blue) and right (red) directions, respectively.     (c) Left ($V_b^l$) and right ($V_b^r$) butterfly velocities' dependence on statistical angle $\theta$ when $U=2$.
Inset: velocity ratio $r= V_b^l/V_b^r$ versus angle $\theta$. (d) Dependence of $V_b^l$ and $V_b^r$ on interaction strength $U$ when $\theta=\pi/2$. Inset: velocity ratio $r=V_b^l/V_b^r$ versus interaction strength $U$.
}
  \label{fig4}
\end{figure}

Figures~\ref{fig4}(a) and (b) further illustrate the OTOC's growth for right and left propagation directions, respectively, with $\theta=\pi/3$ and $U=2$. Indeed, information clearly propagates faster from  right to  left [Fig.~\ref{fig4}(b)] than from  left to right [Fig.~\ref{fig4}(a)]. In order to extract the butterfly velocities most accurately in a finite-size system, we choose the left-most site as the reference point for probing information spreading rightward (and vice-versa for information spreading leftward). We define a butterfly velocity $V_b$ by the boundary of the space-time region where $\vert F_{jk}(t)\vert$ is suppressed by at least 1\% of its initial value.  The linear fits of  butterfly velocities $V_b^{l, r}$ for two directions are shown in the inset of Fig.~\ref{fig4}(b).  The extracted velocities' dependence on $\theta$ and $U$ are further illustrated in Figs.~\ref{fig4}(c) and (d), respectively. As the results show, when $U > 0$ and $0 < \theta < \pi$, the left information propagation velocity is always larger than the right one, with the greatest disparity at intermediate values of $U$ and $\theta$.

{\it Experimental detection.---}%
To study the transport and information dynamics of the AHM, one can experimentally realize the transformed EBHM\@. As mentioned, the correlated-tunneling terms in $\hat{H}_B$ can be engineered using laser-assisted tunneling~\cite{Keilmann11, Greschner15} or lattice shaking~\cite{Strater16, Fan17, Clark18}. Particle transport can be studied  using similar protocols as in previous experiments~\cite{Ronzh13, Schnei12}, where bosonic atoms are first loaded in the center of a 1D optical lattice before being allowed to move under a homogeneous bosonic Hamiltonian. The time-dependent densities, as measured by absorption imaging, directly reflect the anyons' expansion dynamics. On the other hand, measurement of the OTOC defined by Eq.~(\ref{otoc}) is more challenging than mapping out the atomic density. However, instead of measuring \cref{otoc}, one can focus on a bosonic OTOC, $\widetilde{F}_{jk}(t)= \langle \hat{b}_j^{\dagger}(t) \hat{b}_k^{\dagger}(0) \hat{b}_j(t) \hat{b}_k(0)  \rangle$, which, by recent proposals, is experimentally accessible  by inverting the sign of $\hat{H}_B$~\cite{Norman16, Zhu16, Swingle16} or by preparing two identical copies of the system~\cite{Shen17, Bohrdt17}. Numerics show that $\widetilde{F}_{jk}(t)$ can also capture the asymmetric features of OTOC growth~\cite{supp}, thus reflecting anyonic statistics' effect on information dynamics, albeit in an indirect way.

{\it Conclusion and outlook.---}%
We have studied non-equilibrium dynamics of Abelian anyons in a 1D system and found that statistics plays a crucial role in both particle transport and information dynamics.
Our work provides a novel method for detecting anyonic statistics using non-equilibrium dynamics in ultracold atom systems~\cite{Clark18}.

We note the intriguing possibility that a similar dynamical symmetry may exist in other models, such as the $\mathbb{Z}_n$ chiral clock model~\cite{Samajdar18,Seth18}, which has symmetry properties similar to the AHM\@.
Finally, we point out that the inversion symmetry breaking associated with anyonic statistics is also present for non-Abelian anyons in quasi-1D systems~\cite{Clarke13, Cheng12, Mong14}---for example, Majorana fermions (or, more generally, parafermions) at the edge of (fractional) quantum Hall systems, in deep connection with the underlying chirality. We hope this study could motivate future investigation of out-of-equilibrium dynamics and chiral information propagation in these topological systems.

\begin{acknowledgments}
  We thank Chris Flower and Tobias Grass for helpful discussions.
  This work was supported by AFOSR, ARO, NSF PFC at JQI, ARO MURI, ARL CDQI, NSF QIS, NSF Ideas Lab on Quantum Computing, and the DoE ASCR Quantum Testbed Pathfinder program.
  J.R.G. was supported by the NIST NRC Research Postdoctoral Associateship Award.
  D.L.D. acknowledges support from the Laboratory for Physical Sciences,  Microsoft, and the start-up fund from Tsinghua University.
  Z.X.G. acknowledges the start-up fund support from Colorado School of Mines.
  This work was performed in part at the Aspen Center for Physics, which is supported by National Science Foundation grant PHY-1607611.
  The authors acknowledge the University of Maryland supercomputing resources (http://hpcc.umd.edu) made available for conducting the research reported in this paper.
\end{acknowledgments}

\bibliography{anyonbib}

\clearpage
\setcounter{figure}{0}
\makeatletter
\renewcommand{\thefigure}{S\@arabic\c@figure}
\setcounter{equation}{0} \makeatletter
\renewcommand{\thesection}{S.\Roman{section}}
\renewcommand \theequation{S\@arabic\c@equation}
\renewcommand \thetable{S\@arabic\c@table}

\begin{center} 
{\large \bf Supplemental Material}
\end{center}

This Supplemental Material consists of three sections.
In \cref{sec:symmetry}, we derive the dynamical symmetry given by \cref{eq:dynsym1,eq:dynsym2} in the main text.
In \cref{sec:perturb}, we provide an intuitive derivation of the asymmetric expansion dynamics, based on perturbation theory.
In \cref{sec:bosonic_otoc}, we compare features of the bosonic OTOC (which is experimentally accessible) to the anyonic OTOC given by \cref{otoc} in the main text.

\section{Dynamical symmetry of density expansion} \label{sec:symmetry}

In this section, we give detailed derivations for the dynamical symmetry observed in the main text in \cref{eq:dynsym1,eq:dynsym2}.
The inversion symmetry operator $\mathcal{I}$ acts on a bosonic operator as $\mathcal{I} \hat{b}_j \mathcal{I}^\dagger = \hat{b}_{j'}$, where $j'$ is the site that $j$ is mapped to under reflection about the middle of the 1D system.
The time-reversal operator $\mathcal{T}$ acts by complex-conjugating the entries of a state (or operator) written in the bosonic Fock basis; for instance, $\mathcal{T} \hat{b}_j \mathcal{T}^{-1} = \hat{b}_j$ and $\mathcal{T} i \mathcal{T}^{-1} = -i$.
Although $\hat{H}_B$ respects neither time-reversal nor inversion symmetry, it does obey the following $\mathcal{K}$ symmetry~\cite{Lange17}:
\begin{equation}
\mathcal{K} \hat{H}_B \mathcal{K}^{\dagger}= \hat{H}_B,
\label{ksymmetry}
\end{equation}
where $\mathcal{K}= \mathcal{R}\mathcal{I}\mathcal{T}$, and $\mathcal{R}$ is defined as
\begin{equation}
\mathcal{R}= e^{-i\theta \sum_j \hat{n}_j (\hat{n}_j-1)/2}.
\label{rsymmetry}
\end{equation}
With this, we now consider the symmetry properties of the particle dynamics. Using Eq.~(\ref{ksymmetry}), one has:
\begin{equation}
\mathcal{K} e^{-i\hat{H}_Bt} \mathcal{K}^{\dagger} =e^{i\hat{H}_Bt}, 
\label{ksymmetry1}
\end{equation}
where we have used the anti-unitary property of the $\mathcal{K}$ operator.
We first focus on the symmetry properties when flipping the sign of $\theta$ [Eq.\ (\ref{eq:dynsym2}) in the main text]. We label $\hat{H}_B$ with the sign of $\theta$ for convenience:
\begin{equation}
 \hat{H}_{B, \pm \theta}=  -J\sum_{j=1}^{L-1} \( \hat{b}_j^{\dagger} \hat{b}_{j+1}^{\phantom\dagger} e^{\pm i\theta \hat{n}_j} + \Hc \) + \frac{U}{2} \sum_{j=1}^{L} \hat{n}_j (\hat{n}_j-1).
\label{pmtheta}
\end{equation}
The time-dependent density at site $j$ is
\begin{equation}
\langle \hat{n}_j (t)\rangle _{\pm \theta}\equiv \bra{\Psi_0} e^{i \hat{H}_{B, \pm \theta}t }\  \hat{n}_j\  e^{-i \hat{H}_{B, \pm \theta}t } \ket{\Psi_0},
\end{equation}
where $\ket{\Psi_0}$ is the initial Fock product state given in the main text, $\ket{\Psi_0}= \prod_i \hat{b}_i^{\dagger} \ket{0}$. (We have omitted the subscript $``B"$ for simplicity.)
We obtain
\begin{equation} 
\begin{split}
 \ \ \ & \langle \hat{n}_j(t)\rangle_{+\theta}  \equiv \bra{\Psi_0} e^{i \hat{H}_{B, + \theta}t }\  \hat{n}_j\  e^{-i \hat{H}_{B, + \theta}t } \ket{\Psi_0}  \\
  &\quad= \bra{\Psi_0}\mathcal{K}^{\dagger}  e^{-i \hat{H}_{B, + \theta}t } \mathcal{K} \  \hat{n}_j\ \mathcal{K}^{\dagger} e^{i \hat{H}_{B, + \theta}t } \mathcal{K} \ket{\Psi_0} \\
  &\quad= \bra{\Psi_0}  e^{-i \hat{H}_{B, + \theta}t } \mathcal{I} \  \hat{n}_j\ \mathcal{I}^{\dagger} e^{i \hat{H}_{B, + \theta}t } \ket{\Psi_0} \\
 &\quad= \bra{\Psi_0}  e^{-i \hat{H}_{B, + \theta}t } \  \hat{n}_{j^{'}}\  e^{i \hat{H}_{B, + \theta}t } \ket{\Psi_0} ,
\end{split}
\label{split1}
\end{equation}
where, in the second line, we have sandwiched $\mathcal{K}^{\dagger} \mathcal{K}$ between each two operators and used Eq.~(\ref{ksymmetry1}); in the third line,  we have used (i) the fact that when $\mathcal{K}$ operates on the initial state $\ket{\Psi_0}$ in the main text, it gives an unimportant phase after complex conjugation, and (ii) the relation $\mathcal{K} \hat{n}_j \mathcal{K}^{\dagger}= \mathcal{I}\hat{n}_j \mathcal{I}^{\dagger}$; in the fourth line, we have defined the density operator $\hat{n}_{j^{'}}$ on site $j^{'}$, which is related to $\hat{n}_j$ by the inversion symmetry operator $\mathcal{I}$. 

To proceed, we relate $\hat{H}_{B, \pm\theta}$ by the time-reversal symmetry operator $\mathcal{T}$:
\begin{equation}
\mathcal{T} \hat{H}_{B, +\theta} \mathcal{T}^{-1} = \hat{H}_{B, -\theta}. 
\end{equation}
Thus,
\begin{equation}
\mathcal{T} e^{-i \hat{H}_{B, +\theta} t} \mathcal{T}^{-1} = e^{i \hat{H}_{B, -\theta}t}.
\end{equation}
Substituting the above equation into Eq.~(\ref{split1}), we get: 
\begin{equation}
\begin{split}
&\langle \hat{n}_j(t)\rangle_{+\theta}  = 
  \bra{\Psi_0}  e^{-i \hat{H}_{B, + \theta}t } \  \hat{n}_{j^{'}}\  e^{i \hat{H}_{B, + \theta}t } \ket{\Psi_0} \\
  &\quad=  \bra{\Psi_0} \mathcal{T}^{-1}  e^{i \hat{H}_{B, - \theta}t } \mathcal{T} \  \hat{n}_{j^{'}}\ \mathcal{T}^{-1}  e^{-i \hat{H}_{B, - \theta}t } \mathcal{T} \ket{\Psi_0}\\
   &\quad=  \bra{\Psi_0}   e^{i \hat{H}_{B, - \theta}t } \  \hat{n}_{j^{'}}\ e^{-i \hat{H}_{B, - \theta}t }  \ket{\Psi_0}\\
   &\quad\equiv \langle \hat{n}_{j^{'}}(t)\rangle_{-\theta}.
\end{split}
\label{sym_theta}
\end{equation}
Finally, we arrive at a very simple equation [Eq.\ (\ref{eq:dynsym2}) in the main text]: $\langle \hat{n}_j(t)\rangle_{+\theta}= \langle \hat{n}_{j^{'}}(t)\rangle_{-\theta} $. This relation just tells us that when flipping the statistical angle $\theta$, the density expectation values are related by inversion, which agrees with our results in Figs.~\ref{fig1}(f) and (g) in the main text. For $\theta = 0$ or $\pi$, we have $\langle \hat{n}_j(t)\rangle_{0, +\pi}= \langle \hat{n}_{j^{'}}(t)\rangle_{0, -\pi}= \langle \hat{n}_{j^{'}}(t)\rangle_{0,+ \pi}$; that is, for the boson case ($\theta=0$) or the pseudofermion case ($\theta= \pi$), the density expands symmetrically whether or not $U=0$.

There remains another dynamical symmetry [Eq.\ (\ref{eq:dynsym1}) in the main text]: when changing the sign of the interaction $U$, one gets the same behavior as changing the sign of $\theta$, i.e.,\ the two density expansions are related by inversion symmetry. Let us now derive this relation.

Like in Eq.~(\ref{pmtheta}), we label $\hat{H}_B$ with the sign of $U$: 
\begin{equation}
 \hat{H}_{B, \pm U}=  -J\sum_{j=1}^{L} \( \hat{b}_j^{\dagger} \hat{b}_{j+1}^{\phantom\dagger} e^{i  \theta \hat{n}_j} + \Hc \) \pm \frac{U}{2} \sum_{j=1}^{L} \hat{n}_j (\hat{n}_j-1).
 \label{pmU}
\end{equation}
Replacing $\hat{H}_{B, +\theta}$ with $\hat{H}_{B, +U}$ in Eq.~(\ref{split1}), we  get
\begin{equation}
\langle \hat{n}_j(t)\rangle_{+U} = 
  \bra{\Psi_0}  e^{-i \hat{H}_{B, + U}t } \  \hat{n}_{j^{'}}\  e^{i \hat{H}_{B,+ U}t } \ket{\Psi_0}.
 \label{eqpmU}
\end{equation}
Now let us define a number parity operator, $\mathcal{P}= e^{i \pi \sum_r \hat{n}_{2r+1}}$, which measures the parity of total particle number on the odd sites. This operator anti-commutes with the first term of Eq.~(\ref{pmU}), but commutes with the second term. Therefore,
\begin{equation}
\begin{split}
\mathcal{P} \hat{H}_{B, +U} \mathcal{P}^{\dagger} &=  J\sum_{j=1}^{L} \( \hat{b}_j^{\dagger} \hat{b}_{j+1}^{\phantom\dagger} e^{i  \theta \hat{n}_j} + \Hc \) + \frac{U}{2} \sum_{j=1}^{L} \hat{n}_j (\hat{n}_j-1) 
\\ &\equiv -\hat{H}_{B, -U}. 
\end{split}
\end{equation}
Thus,
\begin{equation}
\mathcal{P} e^{-i \hat{H}_{B, +U} t} \mathcal{P}^{\dagger} = e^{i \hat{H}_{B, -U}t}.
\end{equation}
Substituting the above equation into Eq.~(\ref{eqpmU}) results in
\begin{equation}
\begin{split}
&\langle \hat{n}_j(t)\rangle_{+U}  = 
  \bra{\Psi_0}  e^{-i \hat{H}_{B, + U}t } \  \hat{n}_{j^{'}}\  e^{i \hat{H}_{B, + U}t } \ket{\Psi_0} \\
  = &  \bra{\Psi_0} \mathcal{P}^{\dagger}  e^{i \hat{H}_{B, - U}t } \mathcal{P} \  \hat{n}_{j^{'}}\ \mathcal{P}^{\dagger}  e^{-i \hat{H}_{B, - U}t } \mathcal{P} \ket{\Psi_0}\\
   =&  \bra{\Psi_0}   e^{i \hat{H}_{B, - U}t } \  \hat{n}_{j^{'}}\ e^{-i \hat{H}_{B, - U}t }  \ket{\Psi_0}\\
   = & \langle \hat{n}_{j^{'}}(t)\rangle_{-U}. 
\end{split}
\label{sym_U}
\end{equation}
Once again, we arrive at a simple expression [Eq.\ (\ref{eq:dynsym1}) in the main text], 
$\langle \hat{n}_j(t)\rangle_{+U} = \langle \hat{n}_{j^{'}}(t)\rangle_{-U}$, which confirms that by changing the sign of interaction $U$, the density expansion of anyons undergoes an inversion operation. For zero interaction strength, we have $\langle \hat{n}_j(t)\rangle_{U=+0} = \langle \hat{n}_{j^{'}}(t)\rangle_{U=-0} = \langle \hat{n}_{j^{'}}(t)\rangle_{U=+0}$. Therefore, the density expansion of anyons is symmetric when $U=0$, regardless of whether $\theta$ is a multiple of $\pi$.

More generally, it is straightforward to show that the dynamical symmetry relations shown in Eqs.~(\ref{sym_theta}) and (\ref{sym_U}) hold for a class of initial states satisfying $\mathcal{K} \ket{\Psi}= e^{i\phi} (\ket{\Psi})^{*} $ for some $\phi$.

\section{Perturbation analysis of asymmetric expansion} \label{sec:perturb}

In this section, we provide intuition while deriving the asymmetric expansion using perturbation theory. Specifically, we show that the interference between the lowest two order terms in the unitary evolution generally gives rise to asymmetric density expansion dynamics. Once again, we focus on the transformed bosonic Hamiltonian ($\hat{H}_B$) for simplicity. 

Using a Taylor expansion, the unitary time evolution operator can be written as  
\begin{equation}
\mathcal{U}= e^{-i \hat{H}_Bt}= \sum_{n=0}^{\infty} \frac{(-i\hat{H}_Bt)^n}{n!}= 1- i\hat{H}_Bt+ \frac{(i\hat{H}_Bt)^2}{2!}- \cdots.
\end{equation}
We assume the initial state $\ket{\psi_0}$ to be a product state (in Fock space) that is inversion symmetric around the lattice center (i.e.,\ $\mathcal{I}\ket{\psi_0}=\ket{\psi_0}$). The final state after time evolution can be expanded as a sum of product states in Fock space. We consider, as target states, a pair of such product states which are related by inversion symmetry, $\ket{\psi_2}= \mathcal{I} \ket{\psi_1}$, and show that their overlaps with the time-evolved state are different due to the interference of the $k$th and $(k+1)$th order terms in the expansion. 

We denote the matrix element corresponding to the $k$th order term evolving $\ket{\psi_0}$ to $\ket{\psi_1}$ as
\begin{equation}
M_k^{(1)}= \Braket{\psi_1 | \frac{(-i\hat{H}_Bt)^k}{k!} | \psi_0}= \frac{(-it)^k}{k!}A_k ,
\end{equation}
where we have defined $A_k= \braket{\psi_1 | \hat{H}_B^k | \psi_0}$. Similarly, $M_k^{(2)}$ is the matrix element from $\ket{\psi_0}$ to $\ket{\psi_2}$ due to the $k$th order term: 
\begin{equation}
M_k^{(2)}= \Braket{\psi_2 | \frac{(-i\hat{H}_Bt)^k}{k!} | \psi_0}= \frac{(-it)^k}{k!}B_k ,
\end{equation}
where $B_k= \braket{\psi_2 | \hat{H}_B^k | \psi_0}$. 
Using the symmetry properties of the Hamiltonian, we can get: 
\begin{equation}
\begin{split}
B_k&=\bra{\psi_2} \hat{H}_B^k\ket{\psi_0}
=\bra{\psi_1}\mathcal{I}^{\dagger}\hat{H}_B^k \mathcal{I}\ket{\psi_0}\\
&= e^{i(\phi_2-\phi_0)}\bra{\psi_1}\mathcal{I}^{\dagger}\mathcal{R}^{\dagger}\hat{H}_B^k \mathcal{R}\mathcal{I}\ket{\psi_0}\\
&= e^{i(\phi_2-\phi_0)}\bra{\psi_1}(\mathcal{T}\hat{H}_B^k \mathcal{T}^{-1})\ket{\psi_0}
\\&= e^{i(\phi_2-\phi_0)}(\bra{\psi_1}\hat{H}_B^k \ket{\psi_0})^{*}\\
&= e^{i(\phi_2-\phi_0)}A_k^{*} ,
\end{split}
\label{k}
\end{equation}
where in the second line, we have used the symmetry relation between $\ket{\psi_1}$ and $\ket{\psi_2}$ and the fact that $\ket{\psi_0}$ is symmetric under $\mathcal{I}$; in the third line, we extract the phase factor associated with the action of the $\mathcal{R}$ symmetry operator [defined in Eq.~(\ref{rsymmetry})] on states $\ket{\psi_{0, 2}}$: $\mathcal{R}\ket{\psi_{0, 2}}= e^{i \phi_{0, 2}}\ket{\psi_{0, 2}}$; in the fourth line, we have used the symmetry property given by Eq.~(\ref{ksymmetry}); and in the fifth line, we have used the fact that the time-reversal operator acting on $\hat{H}_B$ is equivalent to changing the matrix element to its complex conjugate. 

From here forward, let $k$ be the lowest order for which $M_k^{(1)}$ [or, equivalently, $M_k^{(2)}$] is non-zero.  Because the Hamiltonian $\hat{H}_B$ can have non-zero interactions $U$, the $(k+1)$th expansion terms could also evolve the initial state to $\ket{\psi_{1,2}}$.
Therefore, we consider the leading two order terms which contribute to the matrix element for $\bra{\psi_{1,2}} \mathcal{U} \ket{\psi_0}$: $M_{k}^{(1,2)}$ and $M_{k+1}^{(1,2)}$. We define $S_{1,2}$ to be amplitudes including the total contribution of the $k$th and $(k+1)$th orders: 
\begin{align}
S_1= \left\lvert M_k^{(1)}+ M_{k+1}^{(1)}\right\rvert &= \frac{t^k}{k!} \left\lvert A_k+ \frac{-it}{k+1}A_{k+1}\right\rvert , \label{s_1} \\
S_2=  \left\lvert M_k^{(2)}+ M_{k+1}^{(2)}\right\rvert &= \frac{t^k}{k!} \left\lvert B_k+ \frac{-it}{k+1}B_{k+1}\right\rvert. \label{s_2}
\end{align}
Using \cref{k}, Eq.~(\ref{s_2}) can be re-written as
\begin{equation}
\begin{split}
S_2&=   \frac{t^k}{k!} \left\lvert B_k+ \frac{-it}{k+1}B_{k+1}\right\rvert= \frac{t^k}{k!} \left\lvert A_k^*+ \frac{-it}{k+1}A_{k+1}^*\right\rvert \\
  &= \frac{t^k}{k!} \left\lvert A_k- \frac{-it}{k+1}A_{k+1}\right\rvert.
\end{split}
\label{s_3}
\end{equation}
 Comparing \cref{s_1,s_3}, we can see that because the sign before $A_{k+1}$ is different, the two amplitudes $S_1$ and $S_2$ are in general not equal to each other. This is a simple way of understanding the observed asymmetric expansion in the left and right directions. 
 
The following remarks regarding $S_1$ and $S_2$ are in order: (i) If we set $\theta = 0$ or $\theta=\pi$, the matrix elements $A_k$ and $A_{k+1}$ are both real numbers. In this case, $S_1$ and $S_2$ are exactly equal to each other. This implies that for zero statistical angle $\theta$, the perturbation analysis predicts symmetric density expansion, consistent with our numerics. (ii) On the other hand, for non-zero $\theta$, $A_k$ and $A_{k+1}$ are generally complex numbers, and $S_1$ and $S_2$ are not necessarily equal, therefore predicting asymmetric expansion in general. (iii) When $\theta$ reverses its sign, all the matrix elements change to their complex conjugates, and therefore the values of $S_1$ and $S_2$ are swapped. In this way, the anyons reverse their preferred propagation directions, in agreement with the numerical results. (iv) When the interaction strength $U$ is zero, the matrix element $M_{k+1}^{(1)}$ vanishes, since the Hamiltonian only has hopping terms and hopping once more could not get back to the same state configuration as $\ket{\psi_{1,2}}$. Therefore, $S_1$ and $S_2$ are the same when $U=0$. (v) When $U$'s sign is reversed, $A_{k+1}$ also reverses its sign, therefore swapping the values of $S_1$ and $S_2$.  Thus, the anyons once again reverse their preferred propagation directions.  
 
The above  analysis is completely consistent with the numerical results in the main text. We have once again demonstrated that the crucial ingredients for asymmetric expansion are non-zero statistics $\theta$ and interaction $U$. To illustrate more clearly the above derivations, we consider a very simple example for clarification.  Let us choose $\ket{\psi_0}= \ket{\cdots 0110\cdots } $, $\ket{\psi_1}= \ket{\cdots 0011\cdots } $,  $\ket{\psi_2}= \ket{\cdots 1100\cdots } $. In this case, the second-  and third-order terms in the perturbative time evolution could evolve $\ket{\psi_0}$ to $\ket{\psi_1}$ if $U$ is non-zero. For second-order processes, there are two paths one can start from $\ket{\psi_0}$ and end up with  $\ket{\psi_1}$: either $\ket{\cdots 0110\cdots }\rightarrow \ket{\cdots 0101\cdots } \rightarrow \ket{\cdots 0011\cdots } $ or  $\ket{\cdots 0110\cdots }\rightarrow \ket{\cdots 0020\cdots } \rightarrow \ket{\cdots 0011\cdots } $. The two paths contribute to a total second-order matrix element $\bra{\psi_1}\hat{H}_B^2\ket{\psi_0}= J^2+ J^2e^{i\theta}$.  Due to the on-site interactions, there is also a third-order process which evolves $\ket{\psi_0}$ to $\ket{\psi_1}$:  $\ket{\cdots 0110\cdots }\rightarrow \ket{\cdots 0020\cdots } \rightarrow \ket{\cdots 0020\cdots } \rightarrow \ket{\cdots 0011\cdots } $, whose matrix element is  $\bra{\psi_1}\hat{H}_B^3\ket{\psi_0}=  J^2Ue^{i\theta}$. The total amplitude for second and third order processes is $S_1= \frac{t^2}{2}  \lvert  J^2(1+e^{i\theta})+ \frac{-it}{3} J^2U e^{i\theta}\rvert$. Similarly we can also obtain $S_2= \frac{t^2}{2}  \lvert  J^2(1+e^{-i\theta})+ \frac{-it}{3} J^2U e^{-i\theta}\rvert$. For non-zero $\theta$ and $U$, $S_1 \neq S_2$, implying asymmetric expansion.
The expressions also predict that  the expansion changes its preferred direction when either $\theta$ or $U$ reverses its sign.

\section{Numerical comparison of anyonic and bosonic out-of-time-ordered correlators} \label{sec:bosonic_otoc}

In this section, we provide numerical results for  the bosonic  OTOC, $\widetilde{F}_{jk}(t)= \langle \hat{b}_j^{\dagger}(t) \hat{b}_k^{\dagger}(0) \hat{b}_j(t) \hat{b}_k(0)  \rangle$,  to illustrate that such experimentally measurable quantities can indeed capture the asymmetric information spreading. 

\begin{figure}[b]
  \centering\includegraphics[width=0.5\textwidth]{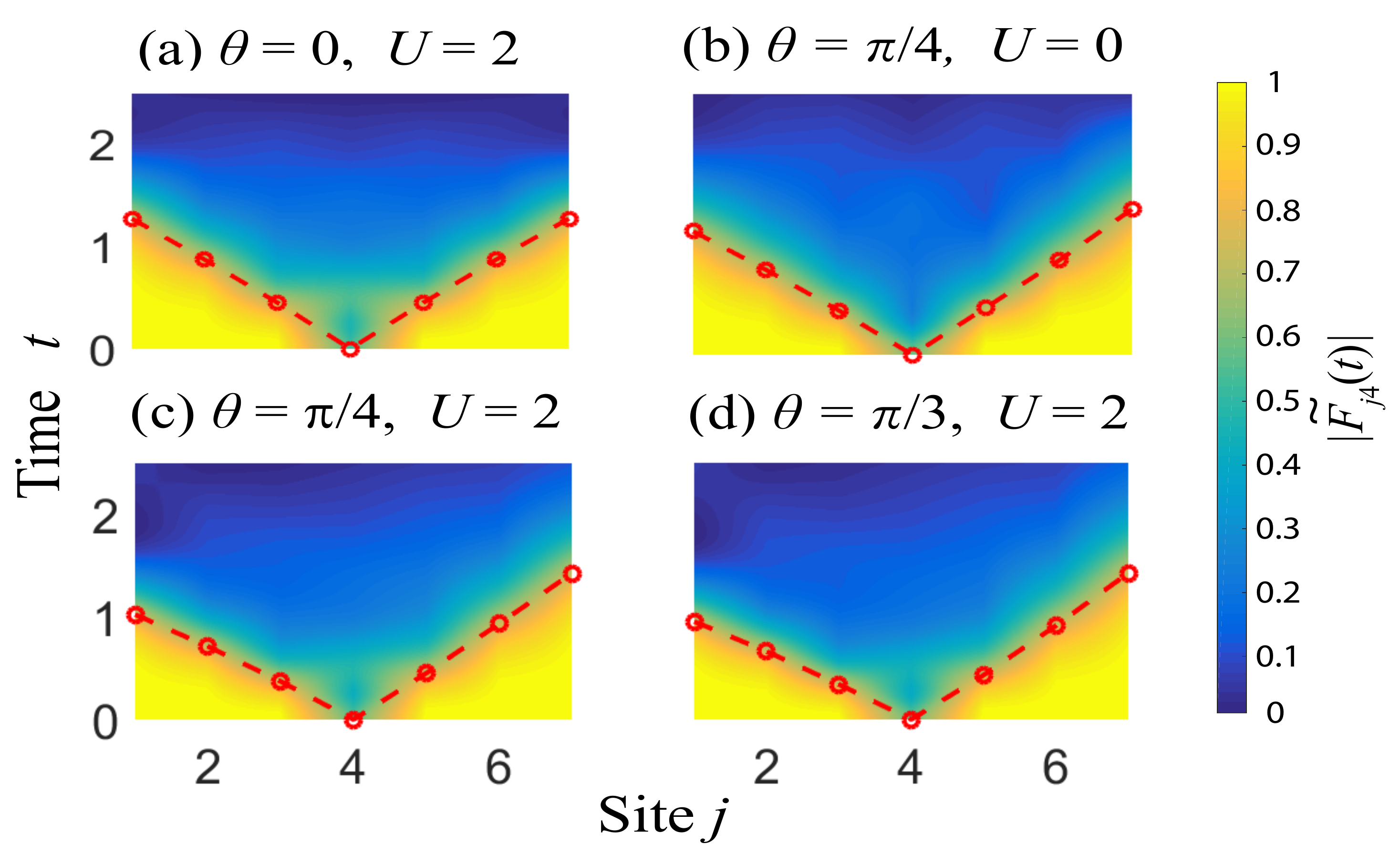}
  \caption{Growth of the bosonic OTOC $|\widetilde{F}_{jk}(t)|$  for different statistical angles $\theta$ and interaction strengths $U$. (a) Bosonic case ($\theta=0$) with interaction strength $U=2$. Anyonic case with (b) vanishing  and (c),(d) non-vanishing  interaction strengths. As in \cref{fig3}, $L=7$, $\beta^{-1}=6$, the local Hilbert space of each site is truncated to three states, and the red dots denote the OTOC falling to 75\% of its initial value.}
  \label{figs1}
\end{figure}

\Cref{figs1} shows the bosonic OTOC growth, with parameters the same as \cref{fig3} in the main text. As one can see, the bosonic OTOCs with non-zero statistical angle also exhibit asymmetric information propagation, similar to their anyonic counterparts.

\begin{figure}[tb]
  \centering\includegraphics[width=0.45\textwidth]{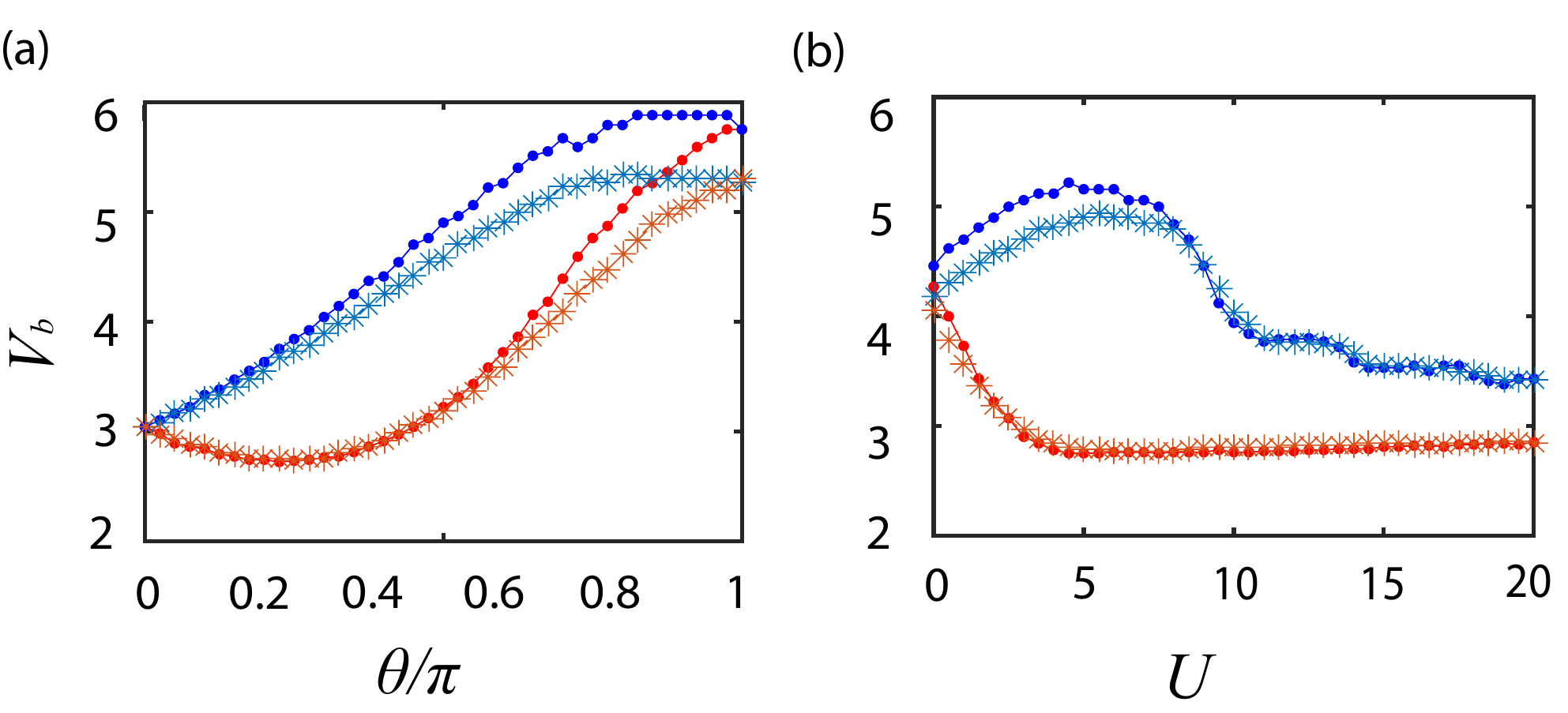}
  \caption{Comparison of butterfly velocities  extracted from the anyonic (dots) and bosonic (asterisks) OTOCs' growth. (a) The butterfly velocities' dependence on statistical angle $\theta$ for fixed $U=2$. The blue dots/asterisks denote the butterfly velocities in the left direction, while the red dots/asterisks denote the butterfly velocities in the right direction.  (b) Similar to (a), but for fixed statistical angle $\theta= \pi/2$  and varying interaction strength $U$.}
  \label{figs2}
\end{figure}

\Cref{figs2} shows the butterfly velocities extracted from the bosonic OTOC\@. In order to make comparisons to anyonic results, we also plot data from Figs.~\ref{fig4}(c) and (d) of the main text. As the figures illustrate, the bosonic butterfly velocities are highly asymmetric for the left and right propagation directions. Moreover, in the regimes of either small $\theta$ or large $U$,  both the left and right velocities of the bosonic OTOC agree well with the anyonic OTOC\@. This can be understood intuitively, as the fractional Jordan-Wigner transformation has reduced effect at small $\theta$, and large $U$ corresponds to the hard-core limit, where anyonic statistics becomes less important.  Moreover, the bosonic/anyonic plots in \cref{figs2} share qualitative features for all values of $\theta$ or $U$.  This suggests that the bosonic OTOC also exhibits signatures of the asymmetric propagation of information due to anyonic statistics. 

\end{document}